\documentclass{ws-procs9x6}            % default, citations in superscript

\usepackage[varg]{txfonts}   % Web of Conferences font
\usepackage{siunitx}
\usepackage{array}
\usepackage{amsmath}
\usepackage{tikz}
\newcommand{\preliminary}[2][3]{
\begin{tikzpicture}
	\node[anchor=south west,inner sep=0] (murks) at (0,0) {#2};
  \begin{scope}[x={(murks.south east)},y={(murks.north west)}]
    \node [rotate=30,scale=2,text opacity=0.2]
    at (murks.center) {preliminary};
  \end{scope}
\end{tikzpicture}
}
\usepackage{float}
\usepackage{wrapfig}
\usepackage{stfloats}

\usepackage[caption=false]{subfig}

\begin{document}
\title{Strangeness photoproduction at the BGO-OD experiment }

% \author{A. B. Author$^*$ and C. D. Author$^1$}

%\author[1]{Alice Smith}
%\author[2]{Bob Jones}
%\affil[1]{Department of Mathematics, University X}
%\affil[2]{Department of Biology, University Y}

\author{Georg Scheluchin$^1$, 
Stefan Alef$^1$, 
Patrick Bauer$^1$, 
Reinhard Beck$^2$, 
Alessandro Braghieri$^{13}$, 
Philip Cole$^3$, 
Rachele Di Salvo$^4$,  
Daniel Elsner$^1$, 
Alessia Fantini$^{4,5}$, 
Oliver Freyermuth$^1$, 
Francesco Ghio$^{6,7}$, 
Anatoly Gridnev$^8$,  
Daniel Hammann$^1$,  
J\"urgen Hannappel$^1$,  
Thomas Jude$^1$, 
Katrin Kohl$^1$, 
Nikolay Kozlenko$^8$, 
Alexander Lapik$^9$,  
Paolo Levi Sandri$^{10}$,  
Valery Lisin$^9$,  
Giuseppe Mandaglio$^{11,12}$, 
Roberto Messi$^{4,5}$, 
Dario Moricciani$^4$, 
Vladimir Nedorezov$^9$,  
Dmitry Novinsky$^8$,  
Paolo Pedroni$^{13}$, 
Andrei Polonski$^9$, 
Bj\"orn-Eric Reitz$^1$, 
Mariia Romaniuk$^4$, 
Hartmut Schmieden$^1$,  
Victorin Sumachev$^8$, 
Viacheslav Tarakanov$^8$,and  
Christian Tillmanns$^1$}

\address{$^1$Rheinische Friedrich-Willhelms-Universit\"at Bonn, Physikalisches Institut, Nu\ss allee 12, 53115 Bonn, Germany \\ 
$^2$Helmholtz-Institut fuer Strahlen- und Kernphysik, Universitaet Bonn, Nussallee 1-16, D-53115 Bonn Germany \\
$^3$Lamar University, Department of Physics, Beaumont, Texas, 77710, USA \\
$^4$INFN Roma ``Tor Vergata", Rome, Italy \\
%INFN Roma Tor Vergata, Via della Ricerca Scientifica 1, 00133 Rome, Italy \\
$^5$Università di Roma ``Tor Vergata", Via della Ricerca Scientifica 1, 00133 Rome, Italy \\
$^6$INFN sezione di Roma La Sapienza, P.le Aldo Moro 2, 00185 Rome, Italy \\
$^7$Istituto Superiore di Sanita, Viale Regina Elena 299, 00161 Rome, Italy \\
$^8$Petersburg Nuclear Physics Institute, Gatchina, Leningrad District, 188300, Russia \\
$^9$Russian Academy of Sciences Institute for Nuclear Research, prospekt 60-letiya Oktyabrya 7a, Moscow 117312, Russia \\
$^{10}$INFN - Laboratori Nazionali di Frascati, Via E. Fermi 40, 00044 Frascati, Italy \\
$^{11}$INFN sezione Catania, 95129 Catania, Italy \\
$^{12}$Universita degli Studi di Messina, Via Consolato del Mare 41, 98121 Messina, Italy\\
$^{13}$INFN sezione di Pavia, Via Agostino Bassi, 6 - 27100 Pavia, Italy}

% 
% \address{University Department, University Name,\\
% City, State ZIP/Zone, Country\\
% $^*$E-mail: ab\_author@university.com\\
% www.university\_name.edu}
% 
% \author{A. N. Author}
% 
% \address{Group, Laboratory, Street,\\
% City, State ZIP/Zone, Country\\
% E-mail: an\_author@laboratory.com}

\begin{abstract}
The BGO-OD experiment at the University of Bonn's ELSA accelerator facility in Germany is ideally suited to investigate photoproduction at extreme forward angles. It combines a highly segmented BGO electromagnetic calorimeter at central angles and an open dipole magnetic spectrometer in the forward direction. This allows the detection of forward going kaons, and complex final states of mixed charge from hyperon decays.

Current projects at the BGO-OD experiment include strangeness production of $\gamma p \rightarrow K^+ \Lambda/\Sigma^0$ at forward angles, $K^0\Sigma^0$ with a deuteron target and $K^+\Lambda(1405)$ line shape and cross section measurements.

\end{abstract}


\bodymatter

%%%%%%%%%%%%%%%%%%%%%%%%%%%%%%%%%%%%%%%%%%%%%%%%%%%%%%%%%%%%%%%%%%%%%%%
%----------------------------------------------------------------------
%%%%%%%%%%%%%%%%%%%%%%%%%%%%%%%%%%%%%%%%%%%%%%%%%%%%%%%%%%%%%%%%%%%%%%%

\section{Introduction}
\label{s_intro}

Hadron spectroscopy is used to investigate the degrees of freedom of the constituents of the nucleon. Since the conception of the quark model, there have been descriptions of baryons and mesons with more than three and two valence quarks respectively. Such hadrons could manifest as penta- and tetraquarks, or as meson-meson and meson-baryon molecular like states. Candidates for such exotic matter were found in recent years in the charm sector \cite{bib_X_state}$^,$\cite{bib_LHCb_state}, and there is evidence that similar configurations may exist in the light, strange sector \cite{bib_k0sigma_oset}. To study such effects in photoproduction experiments, access to a low momentum exchange region, where the meson is produced at forward angles is crucial.

The BGO-OD experiment at the University of Bonn's ELSA accelerator facility in Germany is ideally suited for this endeavor. %The setup is described into more detail in section \ref{s_bgood}.
It combines a highly segmented BGO electromagnetic calorimeter at central angles and an Open Dipole magnetic spectrometer in the forward direction. This allows the detection of forward going kaons, and complex final states of mixed charge from hyperon decays (see section~\ref{s_bgood}).
Preliminary results in current projects are shown in section \ref{s_reconst} for $\Lambda(1405)$ production, in section \ref{s_k0sigma0} for $\gamma n \rightarrow K^0 \Sigma^0$ and in section \ref{s_kforward} cross section measurement of $\gamma p \rightarrow K^+ \Lambda/\Sigma^0$ in the extreme forward regime. The presented results are only a handful of currently pursued projects.

%%%%%%%%%%%%%%%%%%%%%%%%%%%%%%%%%%%%%%%%%%%%%%%%%%%%%%%%%%%%%%%%%%%%%%%
%----------------------------------------------------------------------
%%%%%%%%%%%%%%%%%%%%%%%%%%%%%%%%%%%%%%%%%%%%%%%%%%%%%%%%%%%%%%%%%%%%%%%
\section{The BGO-OD experiment}
\label{s_bgood}

\begin{figure*}[h]
\centering
\includegraphics[width=0.8\textwidth,clip]{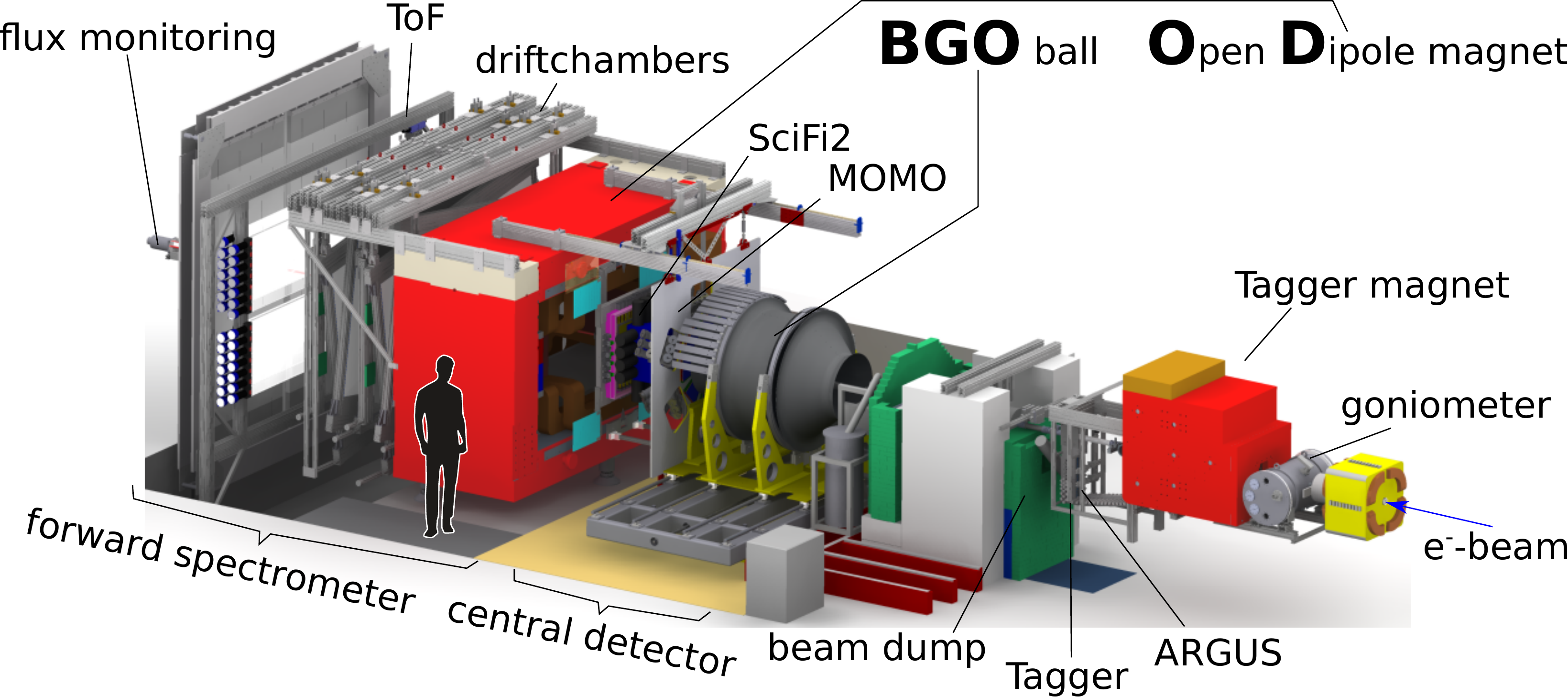}
\caption{Overview of the BGO-OD experiment.}
\label{f_bgood}       % Give a unique label
\end{figure*}

The BGO-OD experiment, shown in figure \ref{f_bgood},  is located at the \textbf{E}lectron \textbf{S}tretcher \textbf{A}ccelerator (ELSA) facility in Bonn. ELSA is an electron accelerator with a quasi continuous beam up to \SI{3.2}{GeV} \cite{bib_hillert}. The electron beam hits a radiator inside the goniometer tank, which creates a real photon beam via bremsstrahlung. The energy of the photons is determined by a measurement of the bremsstrahlung electron with the Tagger detector. The photon beam interacts with the target cell inside the BGO ball, which is filled with liquid hydrogen or deuterium. The final state particles of the reaction are detected with the BGO calorimeter using bismuth germinate oxide crystals between polar angles \SIrange{25}{155}{\degree}. Charged particles traveling in $\theta<$\SI{12}{\degree} are detected in the forward detector. The track trajectory before the open-dipole magnet is measured with the MOMO and SciFi2 scintillating fibre detectors, and the drift chambers measure the trajectory after the magnet. The measured track curvature is used to determine the momentum, while the ToF walls at the end of the experiment measure the velocity via time of flight.

%%%%%%%%%%%%%%%%%%%%%%%%%%%%%%%%%%%%%%%%%%%%%%%%%%%%%%%%%%%%%%%%%%%%%%%
%----------------------------------------------------------------------
%%%%%%%%%%%%%%%%%%%%%%%%%%%%%%%%%%%%%%%%%%%%%%%%%%%%%%%%%%%%%%%%%%%%%%%
\section{$\Lambda(1405)$ photoproduction}
\label{s_reconst}

\begin{figure}[h]
\centering
\subfloat[][Line shape results for $E_\gamma=1500..2300$ MeV. The dashed blue line marks the predictions by J.C.Nacher\cite{bib_Nacher}.]{\preliminary{\includegraphics[width=0.45\textwidth,clip]{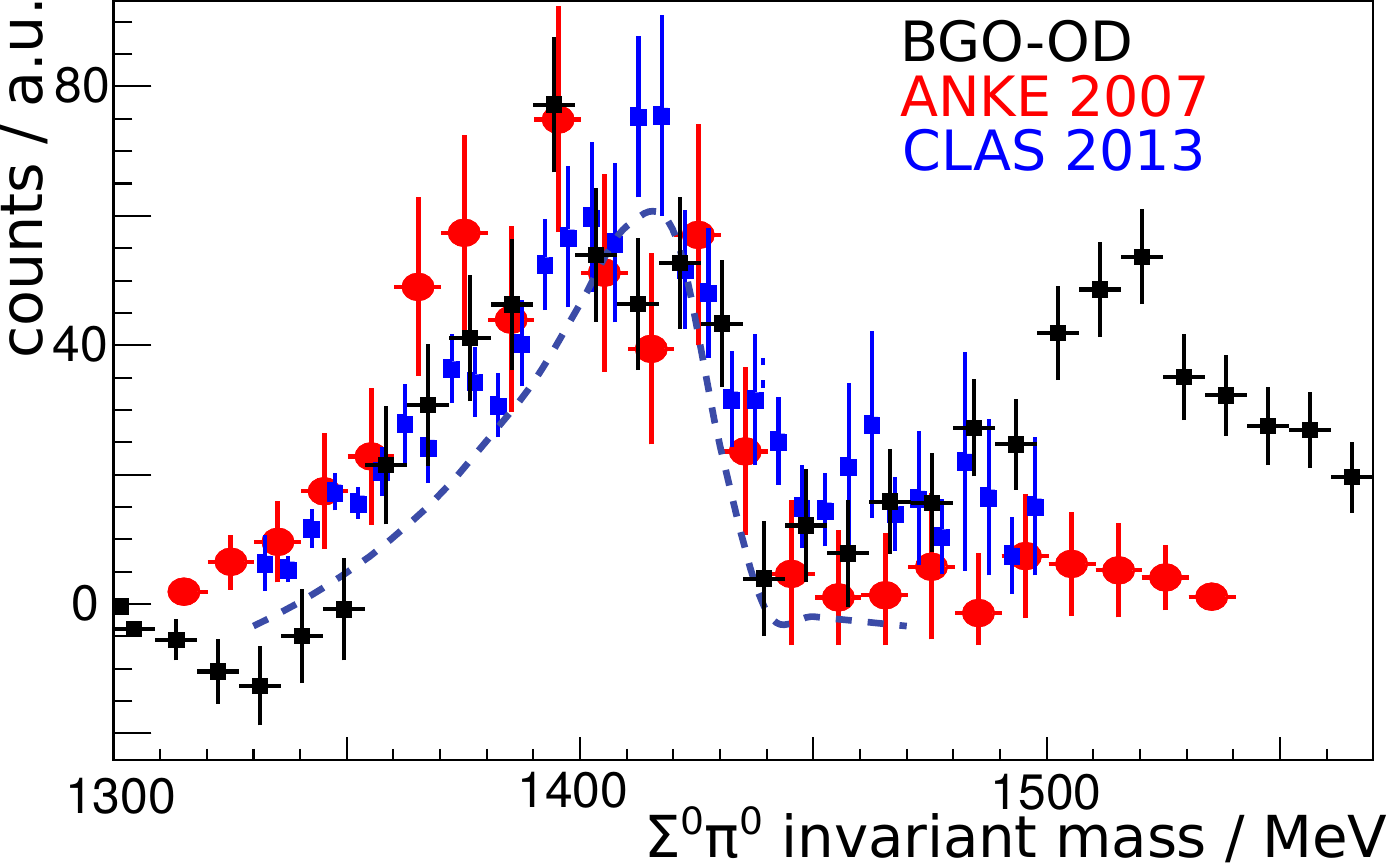}}\label{f_l1405_line}}
\subfloat[][differential cross section]{\preliminary{\includegraphics[width=0.45\textwidth,clip]{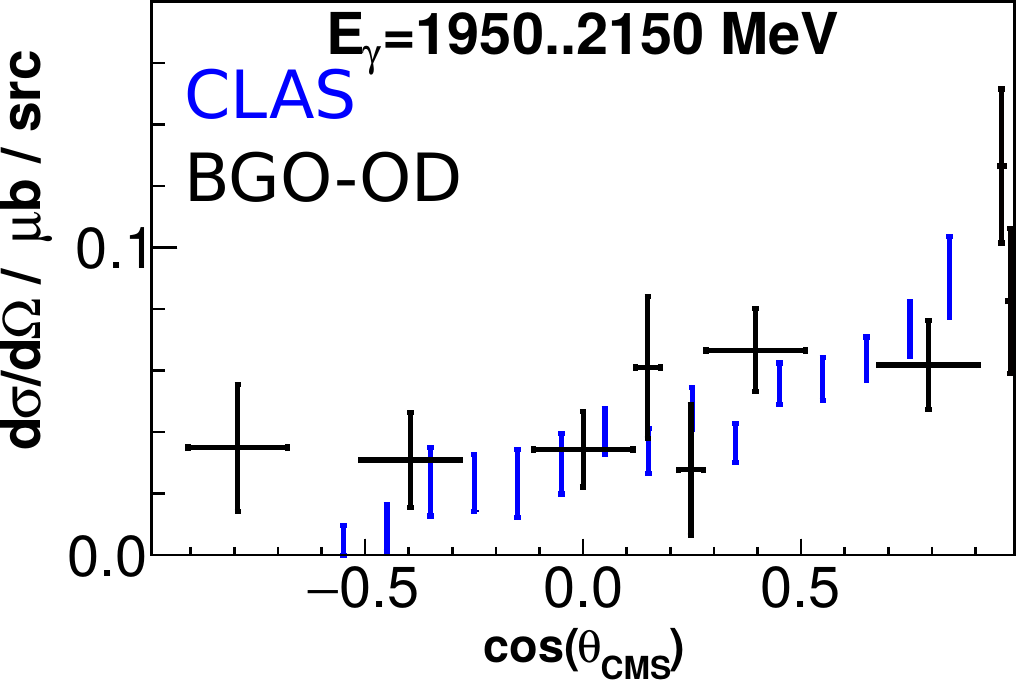}}\label{f_l1405_cross}}
\caption{Preliminary results on $\gamma p \rightarrow K^+ \Lambda(1405) \rightarrow K^+ \pi^0 \Sigma^0$. (a) shows the line shape, while (b) the differential cross section. Black points show the results of the full topology reconstruction and magenta for the $K^+$ in the forward detector. The blue data points were take from the CLAS experiment\cite{bib_CLAS_lineshape}\cite{bib_CLAS_lineshape2} and the red data from the ANKE experiment\cite{bib_anke}.} 
\label{f_l1405}
\end{figure}

Since years it is speculated that the $\Lambda(1405)$ has a molecular like structure of $N\bar{K}$ \cite{Guo:2017jvc}\cite{bib_Nacher}. Lattice QCD calculations give more support for the molecule like structure compared to a genuine three quark state \cite{bib_latticeQCD}\cite{bib_molina}. BGO-OD is ideally suited to detect the $\Lambda(1405)\rightarrow\pi^0\Sigma^0$ decay.
%It is fair to assume that this bounding occurs preferably with low momentum difference between this two particles. In a fixed target experiment with the reaction $\gamma p \rightarrow K^+ \Lambda(1405)$ this would mean kinematics where the $K^+$ takes most of the momentum and consequently extreme forward direction. 
% 
% $\Lambda(1405)$ was identified via the $\Sigma^0\pi^0$ decay, which is prohibited for $\Sigma(1385)$. The complete reaction is $\gamma p \rightarrow K^+ \Lambda(1405) \rightarrow K^+ \Sigma^0 \pi^0$. The reaction can be identified by detecting $K^+$ and $\pi^0$, while the $\Sigma^0$ is identified via missing mass techniques. This can be achieved with the $K^+$ detected in the forward spectrometer. An additional technique is used by detecting the full final state increasing the $K^+$ polar angle acceptance in exchange for lower statistics.  
The preliminary results on the line shape of the $\pi^0\Sigma^0$ decay are compared to other experimental results in figure \ref{f_l1405_line}, while figure \ref{f_l1405_cross} shows an energy bin of the measured differential cross section. The line shape and cross section agrees to the other experimental data within statistics, while the forward spectrometer allows the extension of the CLAS cross section to more forward angles.

% \begin{figure}[h]
% \centering
% \preliminary{
% \includegraphics[width=0.45\textwidth,clip]{LineShape.pdf}}
% \caption{Preliminary line shape of $\Lambda(1405)$ compared to the CLAS experiments in blue\cite{bib_CLAS_lineshape2} and ANKE in red\cite{bib_anke}. The dashed blue line marks the predictions by J.C.Nacher\cite{bib_Nacher}.  \newline $E_\gamma=1500..2300$ MeV. }
% \label{f_lineshapes}       % Give a unique label
% \end{figure}
% 
\newpage

\section{$K^0\Sigma^0$ with deuterium target}
\label{s_k0sigma0}

%\begin{figure}[h]
\begin{wrapfigure}{r}{5.3cm}
\centering
\vspace{-1cm}
\preliminary{\includegraphics[trim={0cm 0cm 0cm, 0cm},clip,width=0.45\textwidth]{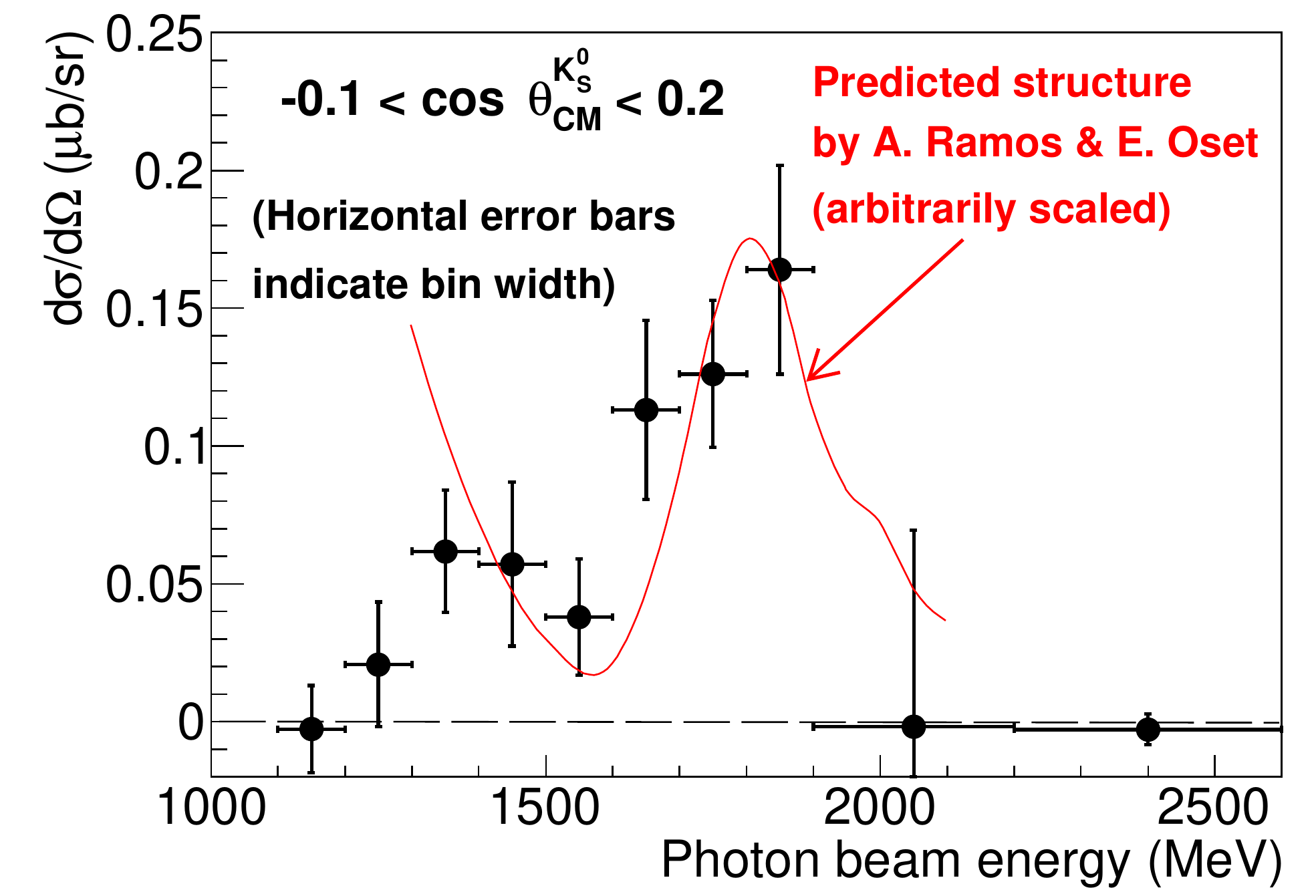}}
\vspace{-0.5cm}
\caption{Preliminary differential cross section on $\gamma n(p)\rightarrow K^0\Sigma^0(p)$ using a deuterium target. The red line shows the predictions for the total cross section \cite{bib_k0sigma_oset}.}
\label{f_crosssection_k0sigma0}       % Give a unique label
%\end{figure}
\end{wrapfigure}

Previous measurements of the \mbox{$\gamma p \rightarrow K^0 \Sigma^+$} cross section showed a cusp over the $K^*Y$ thresholds\cite{bib_CBTABS_k0sigma}. Similar models which predicted the pentaquark at LHCb, predict a rise in the $K^0\Sigma^0$ cross section at the exact same energy\cite{bib_k0sigma_oset}. In these models a $K^*$ is created below threshold, feeding the $K^0\Sigma^+$ cross section. At BGO-OD deuteron data was analyzed on the $\gamma n(p) \rightarrow K^0 \Sigma^0(p)$ reaction by K. Kohl. The preliminary results are shown in figure \ref{f_crosssection_k0sigma0}. They show consistency with the predicted structure of the interaction.

\section{$K^+$ at extreme forward angles}
\label{s_kforward}
\begin{figure}[H]
\centering
\subfloat[][$\gamma p \rightarrow K^+\Lambda$.  Red data shows the \newline  CLAS results of McCracken\cite{bib_kplambda_mccracken}.]{\preliminary{\includegraphics[width=0.46\textwidth]{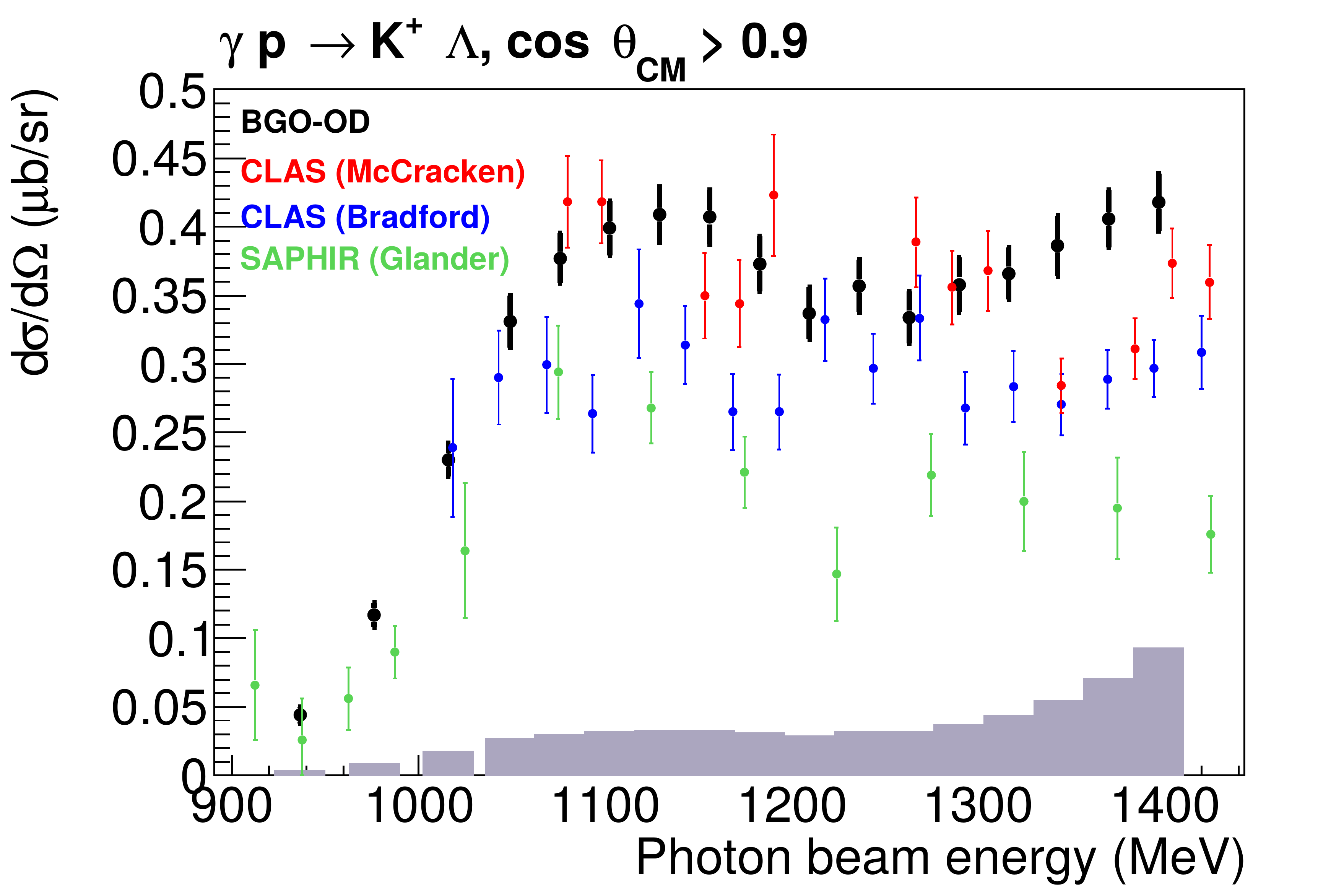}}\label{f_crosssection_kplusforward_lambda}}
\subfloat[][$\gamma p \rightarrow K^+\Sigma^0$.  Red data shows the \newline  CLAS results of Dey\cite{bib_kplambda_dey}.]{\preliminary{\includegraphics[width=0.46\textwidth]{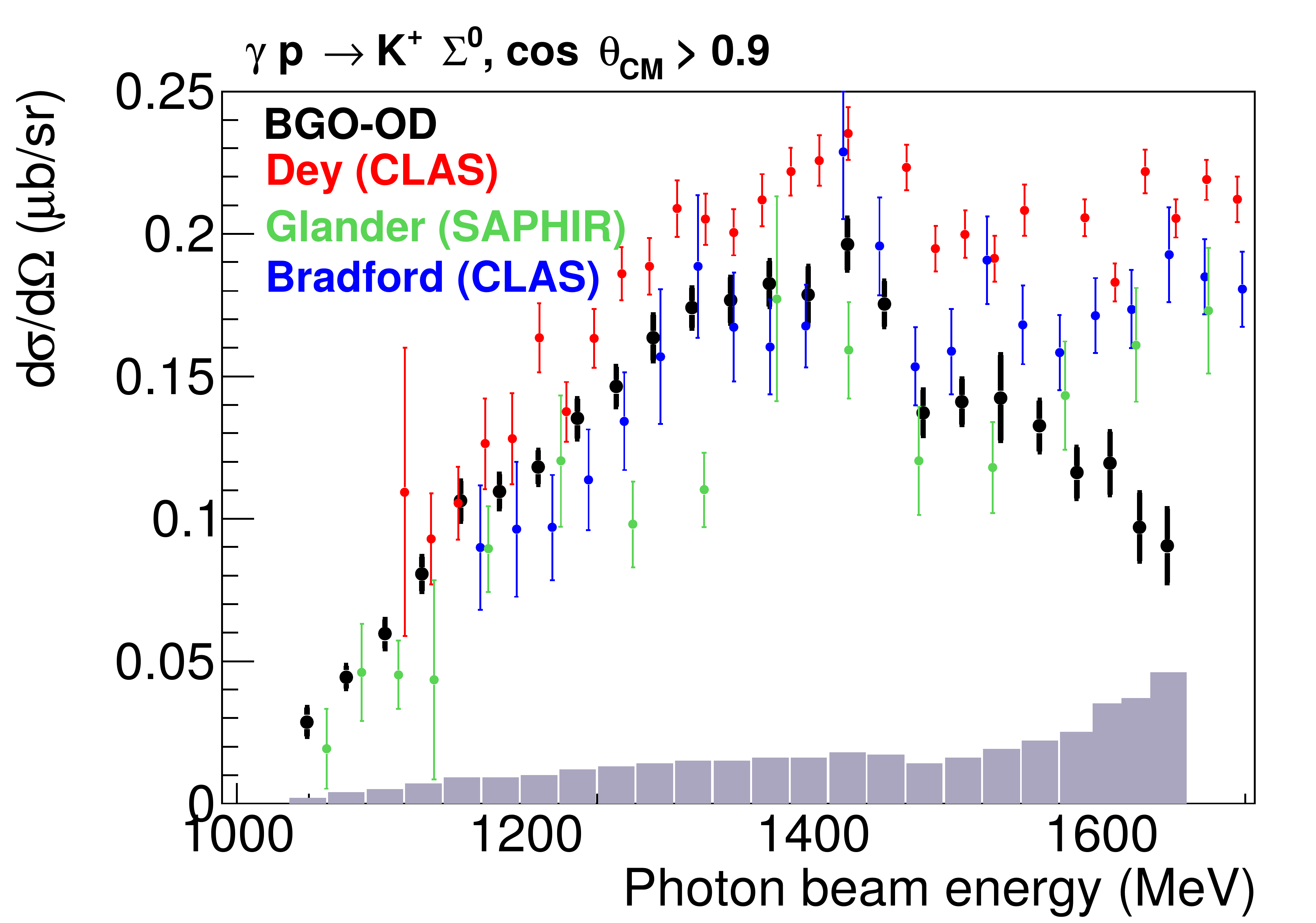}}\label{f_crosssection_kplusforward_sigma}}
\caption{Differential cross section for $\gamma p$ to $K^+\Lambda$ (a) and $K^+\Sigma^0$ (b) for forward going $K^+$. Blue data shows the CLAS results of Bradford\cite{bib_kplambda_bradford}, while green shows the SAPHIR data\cite{bib_kplambda_glander}.} 
\label{f_crosssection_kplusforward}
\end{figure}

The $\gamma p \rightarrow K^+\Lambda$ cross section at forward angles is virtually unconstrained by data, which correspond to low transfer momentum, $t$ \cite{bib_kplambda_models}. Preliminary results on the $K^+\Lambda$ and $K^+\Sigma^0$ cross section in this kinematic regime were analyzed by T.C. Jude at BGO-OD and are shown in figure \ref{f_crosssection_kplusforward}. BGO-OD experiment shows a high statistics compared to other experimental results in the forward region. The high angular resolution of the forward spectrometer allows to map out crucial regions at low $t$. The $K^+\Sigma^0$ cross section shows a apparent cusp close to multiple thresholds, which is still under investigation.

%%%%%%%%%%%%%%%%%%%%%%%%%%%%%%%%%%%%%%%%%%%%%%%%%%%%%%%%%%%%%%%%%%%%%%%
%----------------------------------------------------------------------
%%%%%%%%%%%%%%%%%%%%%%%%%%%%%%%%%%%%%%%%%%%%%%%%%%%%%%%%%%%%%%%%%%%%%%%
\section{Summary and Outlook}
\label{s_summary}

%The BGO-OD experiment is ideally suited to investigate the formation of unconventional states. It combines a highly segmented BGO electromagnetic calorimeter at central angles and an Open Dipole magnetic spectrometer in the forward direction. This allows the detection of forward going kaons with low transfer momentum, and complex final states of mixed charge from hyperon decays. 
Current projects at BGO-OD investigate strangeness photoproduction. The here presented $\gamma p \rightarrow K^+ \Lambda$, $\gamma p \rightarrow K^+ \Sigma^0$, $\gamma p \rightarrow K^+\Lambda(1405)$ and $\gamma n(p) \rightarrow K^0\Sigma^0$ analyses are only a subset of the projects to come. The future projects will include $K^*$ and hypernucleus production. Also non-strange analyses like the $\gamma p \rightarrow \eta'p$ are currently pursued.
%The $K^+\Lambda$ cross section was extended to more forward angles compared to other experiments. The cross sections of $K^+\Sigma^0$ and $K^0\Sigma^0$ show cusps, which could be the result of threshold effects originating from possible meson-baryon states. The $\Lambda(1405)$ is a candidate for such an unconventional state. Line shape measurements with the BGO-OD agree to other experimental data for the $\pi^0\Sigma^0$ decay and other decay channels will be analyzed in the future. 

\subsection*{Acknowledgements}
%I thank the ELSA group for operating and maintaining of the electron accelerator. For the strong support on the maintenance and improvement of our experimental setup I thank the technical staff of the contributing institutions. 

This work is supported by the Deutsche Forschungsgemeinschaft project numbers 388979758 and 405882627, the RSF grant number 19-42-04132, the Third Scientific Committee of the INFN and the European Unions Horizon 2020 research and innovation program under grant agreement No. 824093. 

%
% BibTeX or Biber users please use (the style is already called in the class, ensure that the "woc.bst" style is in your local directory)
% \bibliography{name or your bibliography database}

\begin{thebibliography}{}





\bibitem{bib_X_state}
S.-K. Choi et al. (Belle collaboration), Physical Review Letters \textbf{91}, doi:10.1103/PhysRevLett.91.262001 (2003)
%-----------------------------------------------------------------------
\bibitem{bib_LHCb_state}
R. Aaij et al. (LHCb collaboration), Physical Review Letters \textbf{115}, doi:10.1103/PhysRevLett.115.072001 (2015)
%-----------------------------------------------------------------------
\bibitem{bib_k0sigma_oset}
A.Ramos and E.Oset, Phys. Lett \textbf{B 727}, 287 (2013) 
%-----------------------------------------------------------------------
\bibitem{bib_hillert}
W. Hillert, Europ. Phys. Jour. A \textbf{28}, 139 (2006)
%-----------------------------------------------------------------------
\bibitem{Guo:2017jvc}
 F.~K.~Guo, C.~Hanhart, U.-G.~Meißner, Q.~Wang, Q.~Zhao and B.~S.~Zou,
 %``Hadronic molecules,''
 Rev. Mod. Phys. \textbf{90} (2018) no.1, 015004
 doi:10.1103/RevModPhys.90.015004
%----------------------------------------
%\bibitem{bib-Hemingway}
%R. J. Hemingway, Nucl. Phys. B \textbf{253}, 742 (1985)
%-----------------------------------------------------------------------
\bibitem{bib_Nacher}
J.C.Nacher et al., Physical Review Letters B \textbf{455},55 (1999)
%-----------------------------------------------------------------------
\bibitem{bib_latticeQCD}
J.M.M. Hall et al., Physical Review Letters \textbf{114}, doi:10.1103/PhysRevLett.114.132002 (2015)
%-----------------------------------------------------------------------
\bibitem{bib_molina}
R. Molina and M. Döring, Physical Review Letters D \textbf{94},056010 (2016)
%-----------------------------------------------------------------------
\bibitem{bib_CLAS_lineshape}
K.Moriya et al.(CLAS Collaboration), Physical Review Letters C \textbf{87},035206 (2013)
%-----------------------------------------------------------------------
\bibitem{bib_CLAS_lineshape2}
K.Moriya et al.(CLAS Collaboration), Physical Review Letters C \textbf{88},045201 (2013)

%_----------------------------------------------------------------------
%\bibitem{bib_elsa}
%On basis of website. url: http://www-elsa.physik.uni-bonn.de/index\_en.html 
%-----------------------------------------------------------------------
%\bibitem{bib_cb}
%U. Thoma et al. (CB-ELSA Collaboration), Hirschegg 2001, Structure of hadrons, 193.   (2001)
%-----------------------------------------------------------------------
\bibitem{bib_anke}
I.Zychor et al., Physical Review Letters B \textbf{660},167-171 (2008)
%-----------------------------------------------------------------------
%\bibitem{bib_doublepeak}
%T. Hyodo and D. Jido, Prog.Part.Nucl.Phys. \textbf{67}, 55-98 (2011) 
%-----------------------------------------------------------------------
\bibitem{bib_CBTABS_k0sigma}
R.Ewald et al (CBELSA/TAPS Collaboration), Phys. Lett. \textbf{B 713}, 180 (2012) 
%-----------------------------------------------------------------------
\bibitem{bib_kplambda_models}
D.Skoupil and P.Bydzovsky, Phys. Rev. \textbf{C97}, 025202 (2018)
%-----------------------------------------------------------------------
\bibitem{bib_kplambda_mccracken}
M.E. McCracken et al (CLAS), Phys. Rev. \textbf{C81}, 025201 (2010) 
%-----------------------------------------------------------------------
\bibitem{bib_kplambda_dey}
B.Dey et al. (CLAS), Phys. Rev. \textbf{C82}, 025202 (2010) 
%-----------------------------------------------------------------------
%\bibitem{bib_kplambda_mami}
%T.C. Jude et al (MAMI), Phys. Lett. \textbf{B 735}, 112 (2014)
%-----------------------------------------------------------------------
\bibitem{bib_kplambda_bradford}
R.Bradford et al. (CLAS), Phys. Rev. \textbf{C73}, 035202 (2006)
%-----------------------------------------------------------------------
\bibitem{bib_kplambda_glander}
K.H. Glander et al (SAPHIR), Eur. Phys. J. \textbf{A19}, 251 (2004) 
%-----------------------------------------------------------------------



\end{thebibliography}

\end{document}